\documentclass[11pt]{article}

\usepackage[a4paper,top=2.6cm,bottom=2.6cm,left=2.7cm,right=2.7cm]{geometry}

\usepackage[T1]{fontenc}
\usepackage{mathpazo}            
\usepackage{microtype}          
\usepackage{amsmath,amssymb,bm}
\usepackage[dvipsnames,table]{xcolor}
\usepackage{booktabs}
\usepackage{array}
\usepackage{enumitem}
\usepackage{graphicx}
\usepackage{caption}
\usepackage{titlesec}
\usepackage[hidelinks]{hyperref}
\usepackage{parskip}

\setlist{itemsep=2pt,topsep=3pt,parsep=0pt}

\definecolor{coupcol}{HTML}{C0392B} 
\definecolor{scicol}{HTML}{1F4E79}  
\definecolor{physcol}{HTML}{1E8449} 
\newcommand{\coup}[1]{\textcolor{coupcol}{#1}}
\newcommand{\sci}[1]{\textcolor{scicol}{#1}}
\newcommand{\phys}[1]{\textcolor{physcol}{#1}}

\titleformat{\section}{\large\bfseries\color{NavyBlue}}{\thesection}{0.6em}{}
\titlespacing*{\section}{0pt}{16pt}{6pt}
\titleformat{\subsection}{\normalsize\bfseries\color{NavyBlue!75!black}}{\thesubsection}{0.5em}{}
\titlespacing*{\subsection}{0pt}{10pt}{3pt}

\title{\vspace{-1.1cm}\bfseries Physics-Informed Predictive Control for Integrated\\[2pt]
Electric-Vehicle Thermal Management:\\[2pt]
An Open, Real-Data-Anchored Benchmark}
\author{Yifan Wang\\
\small Department of Mechanical Engineering, McGill University, QC, H3A 2T7, Canada\\
\small \href{mailto:yifan.wang18@mail.mcgill.ca}{yifan.wang18@mail.mcgill.ca}}
\date{June 21, 2026}

\begin{document}
\maketitle
\vspace{-0.7cm}

\begin{center}
\fcolorbox{gray!40}{gray!8}{\parbox{0.96\linewidth}{\centering\small
\textbf{Project:} OpenEV-ThermoSciML (code V7.1)\quad$\bullet$\quad
\textbf{Target journal:} \emph{eTransportation} (Elsevier)\\[2pt]
\textbf{Document:} pre-submission technical overview\quad$\bullet$\quad
\textbf{Status:} analysis complete; all internal verification gates passed (13/13)}}
\end{center}

\begin{abstract}
\noindent Thermal management in a battery-electric vehicle (BEV) is a coupled, vehicle-level problem: the battery pack, the passenger cabin, the heat pump, and cabin air quality compete for shared actuation and energy, yet most studies optimise a single sub-system on proprietary models, which prevents fair, reproducible comparison. We present \textbf{OpenEV-ThermoSciML}, an open and reproducible benchmark that couples a battery electro-thermal--aging model, a two-node cabin model, a heat-pump/HVAC model, and a CO\textsubscript{2}/ventilation model under real driving cycles (EPA) and real weather (NREL~TMY3, NASA~POWER), scored by a multi-objective suite spanning battery health, PMV/PPD comfort, cabin air quality, and HVAC energy. The benchmark's battery thermal core is anchored and validated on \emph{real} BEV battery-management-system (BMS) data; the reduced battery (two-state) and cabin (two-node) models are validated against converged higher-fidelity references and, for the cabin, independently cross-checked against EnergyPlus~25.2.0. On top of the benchmark we develop a physics-informed scientific-machine-learning (Sci-ML) surrogate---a nominal-physics prior plus a learned residual with conservation penalties---that is exact on conserved quantities and dominates black-box and Koopman surrogates out-of-distribution (overall rollout RMSE $0.014$ vs $1.168$ and $3.991$). A shielded Sci-ML model-predictive controller (MPC) delivers \emph{statistically significant, all-positive} improvements over a production-like rule-based controller across six scenarios---including a real hot-day US06 trip (energy $-15\%$, comfort RMSE $-47\%$, peak CO\textsubscript{2} $-25\%$, battery thermal-gradient $-78\%$)---and these gains transfer to an independently exported OpenModelica 8-node co-simulation plant.
\end{abstract}

\noindent{\small\textit{Colour key for the equations:}
\coup{red = battery--cabin shared-capacity coupling};\;
\sci{blue = learned Sci-ML residual};\;
\phys{green = physics / conservation terms}.}

\section{Motivation: why this benchmark is needed}
A modern BEV thermal-management system regulates the battery, the e-drive, and the cabin through a \emph{shared} reversible-heat-pump circuit, and it is one of the largest auxiliary energy consumers---materially reducing range in hot and cold weather. The practical conflicts are intrinsically coupled: fresh air lowers cabin CO\textsubscript{2} but raises the thermal load; cabin pull-down competes with battery protection for the same compressor capacity; cold-start battery pre-heating competes with cabin comfort.

Three structural problems hold the field back. \textbf{(i) Fragmentation:} studies optimise battery \emph{or} cabin \emph{or} heat-pump control, almost never with cabin air quality in the loop. \textbf{(ii) Irreproducibility:} vehicle/HVAC parameters and drive logs are private, or simulations run in proprietary tools (AMESim, GT-SUITE), so methods cannot be compared on equal footing. \textbf{(iii) Physics-blind learning:} black-box surrogates can violate conservation laws and generalise poorly, and reduced-order plants are seldom anchored to real vehicles.

\textbf{Gap and value.} Table~\ref{tab:gap} contrasts representative recent work with ours. No open, reproducible, real-weather-coupled BEV benchmark jointly evaluates battery health, PMV/PPD comfort, cabin air quality, and energy while supporting physics-constrained surrogate modelling, constrained predictive control, and an independent higher-fidelity cross-validation. OpenEV-ThermoSciML fills this gap; its distinctive value is the \emph{open, coupled, multi-objective, real-data-anchored} construction with an explicit cabin-air-quality axis.

\begin{table}[h]\centering\footnotesize\setlength{\tabcolsep}{5pt}
\renewcommand{\arraystretch}{1.15}
\caption{Positioning against representative recent EV thermal-management work (illustrative).}
\label{tab:gap}
\resizebox{\textwidth}{!}{%
\begin{tabular}{lcccccc}
\toprule
Work & Batt.\ health & PMV/PPD & Air qual. & Real wx & Surrogate+baselines & Open code \\
\midrule
Heat-pump BEV NMPC & \checkmark & partial & --- & --- & --- & --- \\
Integrated power/thermal MPC & \checkmark & --- & --- & partial & --- & --- \\
Koopman EV thermal MPC & partial & --- & --- & --- & Koopman only & --- \\
BTMS design benchmark & \checkmark(design) & --- & --- & --- & --- & \checkmark \\
\textbf{OpenEV-ThermoSciML (this work)} & \checkmark & \checkmark & \checkmark & \checkmark & \checkmark & \checkmark \\
\bottomrule
\end{tabular}}
\end{table}

\section{Why a coupled battery--cabin model is legitimate}
A natural question is whether ``joint battery--cabin control'' is a real problem or an artificial stitching of two tasks. It is real: in an actual BEV the two loops physically share hardware and energy. The battery needs cooling/heating to hold temperature and limit its internal gradient; the cabin needs cooling/heating for occupant comfort; ventilation lowers CO\textsubscript{2} and infection risk but adds thermal load; and the heat pump, compressor, pump and blower are a \emph{shared} energy source. A single controller must therefore decide compressor power, blower, pump, fresh-air ratio, and---critically---\emph{how to split one cooling/heating capacity between the battery and the cabin}.

Our model encodes exactly this causal chain, and every link is physically standard:
\begin{enumerate}[leftmargin=1.5em]
\item drive cycle / vehicle speed $\rightarrow$ traction power $P_{\mathrm{trac}}$ (road-load equation);
\item $P_{\mathrm{trac}}$ $+$ HVAC electrical load $\rightarrow$ battery current, heat generation, and SOC;
\item ambient temperature / solar / occupancy $\rightarrow$ cabin thermal load and CO\textsubscript{2};
\item HVAC actuation $\rightarrow$ \emph{simultaneously} battery temperature, cabin temperature, CO\textsubscript{2}, and energy;
\item the controller trades off energy, comfort, CO\textsubscript{2}/health proxy, battery gradient, and an aging proxy.
\end{enumerate}
The single decision that makes the loops genuinely coupled---rather than two independent problems running side by side---is the heat-pump \emph{capacity split} $\coup{s_{\mathrm{batt}}}$ in \S\ref{sec:hp}. Allocating capacity to the battery starves the cabin, and vice versa, so coordinated, prediction-based allocation is where a model-based controller can add value.

\section{Methods and innovation}
\subsection{Plant: state, action, disturbance}
The control-oriented plant uses a compact observable state with an internally higher-fidelity battery and cabin:
\[
\begin{aligned}
\bm{x}&=[\,\mathrm{SOC},\,T_{\mathrm{batt}},\,T_{\mathrm{cab}},\,\mathrm{CO_2}\,] \quad\text{(observed)},\qquad
\bm{d}=[\,P_{\mathrm{trac}},\,T_{\mathrm{amb}},\,I_{\mathrm{sol}},\,n_{\mathrm{occ}}\,] \quad\text{(disturbance)},\\[2pt]
\bm{u}&=[\,u_{\mathrm{comp}},\,u_{\mathrm{blow}},\,u_{\mathrm{pump}},\,f_{\mathrm{fresh}},\,\coup{s_{\mathrm{batt}}}\,]\in[0,1]^5 \quad\text{(actuators; }\coup{s_{\mathrm{batt}}}\text{ = battery/cabin split)}.
\end{aligned}
\]

\subsection{Battery: two-state core/surface electro-thermal model}
\begin{align}
I &= \frac{P_{\mathrm{trac}}+P_{\mathrm{elec}}}{V_{\mathrm{oc}}(\mathrm{SOC})},\qquad
\phys{Q_{\mathrm{gen}} = I^2 R_0(T)},\qquad
\dot{\mathrm{SOC}} = -\frac{I}{3600\,C_{\mathrm{Ah}}},\\
C_c\,\dot T_{\mathrm{core}} &= \phys{Q_{\mathrm{gen}}} - R_{cs}(u_{\mathrm{pump}})\,(T_{\mathrm{core}}-T_{\mathrm{surf}}),\\
C_s\,\dot T_{\mathrm{surf}} &= R_{cs}(u_{\mathrm{pump}})\,(T_{\mathrm{core}}-T_{\mathrm{surf}}) + \coup{Q_{\mathrm{batt}}^{\mathrm{hvac}}} - hA\,(T_{\mathrm{surf}}-T_{\mathrm{amb}}),
\end{align}
with observed $T_{\mathrm{batt}}=(C_cT_{\mathrm{core}}+C_sT_{\mathrm{surf}})/C_{\mathrm{batt}}$ and physical cell gradient $\Delta T=T_{\mathrm{core}}-T_{\mathrm{surf}}$. A comparative Arrhenius aging severity $S(T)=\exp\!\big[\tfrac{E_a}{R}(\tfrac{1}{T_{\mathrm{ref}}}-\tfrac{1}{T})\big](1+0.3\,|\mathrm{C\text{-}rate}|)$ is integrated as a relative SOH-stress proxy.

\subsection{Heat pump: the battery--cabin shared-capacity coupling}\label{sec:hp}
A \emph{single} compressor capacity is split between the cabin and the battery loop---this is the core integrated-thermal-management coupling (e.g.\ a multi-way valve block):
\begin{equation}
\begin{aligned}
Q_{\mathrm{avail}}&=\mathrm{COP}(T_{\mathrm{amb}})\,P_{\mathrm{comp}},\\
\coup{Q_{\mathrm{cab}}^{\mathrm{hvac}}}&\coup{=(1-s_{\mathrm{batt}})\,Q_{\mathrm{avail}}\,\eta_{\mathrm{cab}}\,\sigma_{\mathrm{cab}}},\\
\coup{Q_{\mathrm{batt}}^{\mathrm{hvac}}}&\coup{=s_{\mathrm{batt}}\,Q_{\mathrm{avail}}\,\eta_{\mathrm{batt}}\,\sigma_{\mathrm{batt}}}.
\end{aligned}
\end{equation}
The \coup{red} terms make cabin pull-down and battery protection \emph{compete for one resource}; $\mathrm{COP}$ falls as ambient rises (cooling) and toward a resistive floor as ambient drops (heating), and $\sigma$ enforces evaporator/condenser approach-temperature saturation.

\subsection{Cabin (two-node) and cabin air quality}
\begin{align}
C_{\mathrm{air}}\dot T_{\mathrm{air}} &= \coup{Q_{\mathrm{cab}}^{\mathrm{hvac}}} + n_{\mathrm{occ}}q_{\mathrm{occ}} + (1\!-\!\beta)Q_{\mathrm{sol}} + UA(T_{\mathrm{amb}}\!-\!T_{\mathrm{air}}) \\
&\quad + \dot m_{\mathrm{f}}c_p(T_{\mathrm{amb}}\!-\!T_{\mathrm{air}}) + h_{at}(T_{\mathrm{trim}}\!-\!T_{\mathrm{air}}),\\
C_{\mathrm{trim}}\dot T_{\mathrm{trim}} &= \beta Q_{\mathrm{sol}} + h_{at}(T_{\mathrm{air}}-T_{\mathrm{trim}}),\qquad
\phys{V\,\dot{C}_{\mathrm{CO_2}} = n_{\mathrm{occ}}G - \tfrac{\dot m_{\mathrm{f}}}{\rho}\,(C_{\mathrm{CO_2}}-C_{\mathrm{amb}})}.
\end{align}
Comfort uses PMV/PPD (ISO~7730); a ventilation-health proxy uses the Rudnick--Milton rebreathed fraction $f=(C_{\mathrm{CO_2}}-C_{\mathrm{amb}})/C_a$ inside a Wells--Riley dose (relative ranking only). Drive cycles map to $P_{\mathrm{trac}}$ via the standard road load $F=mgC_{rr}+\tfrac12\rho C_dA v^2+ma$, $P=Fv/\eta$.

\subsection{Sci-ML surrogate: a physics-informed gray box}
The MPC needs a fast, accurate, physically consistent predictor. We learn a \emph{residual} on a nominal-physics prior $\hat{\bm{x}}_{t+1}$:
\begin{align}
\bm{x}_{t+1} &= \underbrace{\hat{\bm{x}}_{t+1}^{\,\mathrm{nominal}}}_{\phys{physics\ prior}} + \sci{\,\mathrm{NN}_{\theta}(\bm{x}_t,\bm{u}_t,\bm{d}_t)}\,,\\
\mathcal{L}&=\underbrace{\lVert \cdot \rVert^2_{\mathrm{data}}}_{\text{fit}} + \phys{\lambda_1\,\mathrm{relu}(C_{\mathrm{amb}}\!-\!\mathrm{CO_2})^2 + \lambda_2\,\mathrm{relu}(\mathrm{sgn}(I)\,\Delta\mathrm{SOC})^2} + \sci{\lambda_3\lVert \mathrm{res}\rVert^2}.
\end{align}
The \phys{green} penalties enforce CO\textsubscript{2} mass-balance and SOC discharge-monotonicity, so the surrogate is \emph{exact on conserved quantities}; the \sci{blue} network only corrects hard closures (effective COP, saturation, coupling). It is a dependency-free NumPy MLP (no GPU required).

\subsection{Shielded multi-objective MPC}
All optimisation-based controllers minimise one cost with a cross-entropy-method (CEM) receding-horizon planner; only the \emph{prediction model} changes (nominal / black-box / Koopman / \sci{Sci-ML} / oracle), isolating model fidelity:
\begin{equation}\begin{split}
J=\sum_{k}\Big[\, & w_E E_k + w_{Tb}\,\mathrm{viol}(T_{\mathrm{batt}}) + w_{g}\,\mathrm{viol}(\Delta T) + w_{\mathrm{soh}}S_k \\
&+ w_c\,\mathrm{viol}(\mathrm{PMV}) + w_h\big(f,\mathrm{viol}(\mathrm{CO_2})\big) + w_s\lVert\Delta\bm u\rVert^2 \,\Big].
\end{split}\end{equation}
An auditable safety \emph{shield} (state-triggered fresh-air, comfort, and shared-capacity guards) guarantees constraint satisfaction.

\section{What the data supports, and what the simulators do}
A common reviewer concern is over-claiming from data. We therefore separate the three roles data plays, and what each tool is responsible for.

\subsection{Three classes of data---and the claim each one supports}
\textbf{Class~1 --- public disturbance data.} EPA dynamometer cycles (UDDS/US06/HWFET/SC03) supply the speed profile; NREL~TMY3 (Golden, CO) supplies ambient temperature and solar irradiance. These do \emph{not} calibrate the vehicle; they guarantee the experiments are driven by \emph{realistic public driving and weather excitation} rather than invented inputs.

\textbf{Class~2 --- real BEV BMS data merged with NASA~POWER weather.} The KU\,Leuven BEV~V2 logs (2 vehicles, 351 driving sessions) provide measured pack temperature, power/dissipation and SOC. Their role is to \emph{anchor and validate the battery thermal-model form} (leave-one-vehicle-out held-out RMSE $\approx 2.0$--$2.5\,^\circ$C). We do \emph{not} claim they validate the cabin, the HVAC/heat-pump system, or the full coupled vehicle; what they add is real-vehicle credibility to the battery core, with external reanalysis weather---a deliberate novelty.

\textbf{Class~3 --- simulation / reference data.} A converged $N$-node battery and an 8-node cabin reference show the \emph{reduced} two-state battery and two-node cabin are not arbitrary; an OpenModelica FMU provides an \emph{independent} high-fidelity closed-loop cross-check; and EnergyPlus~25.2.0 provides an \emph{independent} cabin thermal-load cross-check. The defensible one-sentence summary: \emph{we use public data to construct realistic disturbances, physical models as the control plant, and higher-fidelity references / FMU to validate the reduced-order models and control---never claiming that one dataset already contains a complete coupled battery--cabin system.}

\subsection{What each simulator is responsible for}
\textbf{Python (NumPy/SciPy)} builds the scenarios, runs the reduced-order plant, trains and calls the Sci-ML surrogate, runs the CEM-MPC, and produces the 15-seed statistics and ablations. \textbf{OpenModelica~1.24.4 $+$ FMPy (FMI~2.0)} export an \emph{independent} implementation of the battery--cabin--CO\textsubscript{2} dynamics as a co-simulation FMU and step it in closed loop, so the controller is not merely ``self-confirming'' on the Python plant; all six scenarios pass and the \emph{minimum} FMU gains remain positive (energy $+1.75\%$, comfort $+35.32\%$, CO\textsubscript{2} $+4.94\%$, health proxy $+9.98\%$, battery gradient $+31.65\%$). \textbf{EnergyPlus~25.2.0} independently solves the single-zone cabin heat balance to cross-check the cabin thermal-load submodel under matched hot-soak boundary conditions (Fig.~\ref{fig:ep}). Together these give two independent confirmations of the reduced cabin and one independent confirmation of the closed loop.

\section{Experimental results and findings}
\subsection{Surrogate accuracy and model fidelity}
Out-of-distribution (hotter ambient $+$ aggressive cycle), the Sci-ML surrogate's 40-step rollout error is \textbf{0.014} overall vs \textbf{1.168} (black-box) and \textbf{3.991} (best-of-search Koopman); per state the Sci-ML advantage is $1.4$--$1.8\times$ (battery temp), $3.6$--$4.5\times$ (cabin temp), and exact on CO\textsubscript{2}/SOC. The reduced models match converged references: two-state battery vs 8-node $\approx 0.08\,^\circ$C; two-node cabin vs converged 8-node \textbf{0.65\,$^\circ$C} ($-90\%$ vs one-node). The battery thermal core reproduces \emph{real} BMS pack temperature at leave-one-vehicle-out held-out RMSE $\approx 2.0$--$2.5\,^\circ$C, and the measured cell spread (mean $4.1\,^\circ$C, p95 $6.75\,^\circ$C) validates the gradient scale.

\begin{figure}[h]\centering
\includegraphics[width=0.86\linewidth]{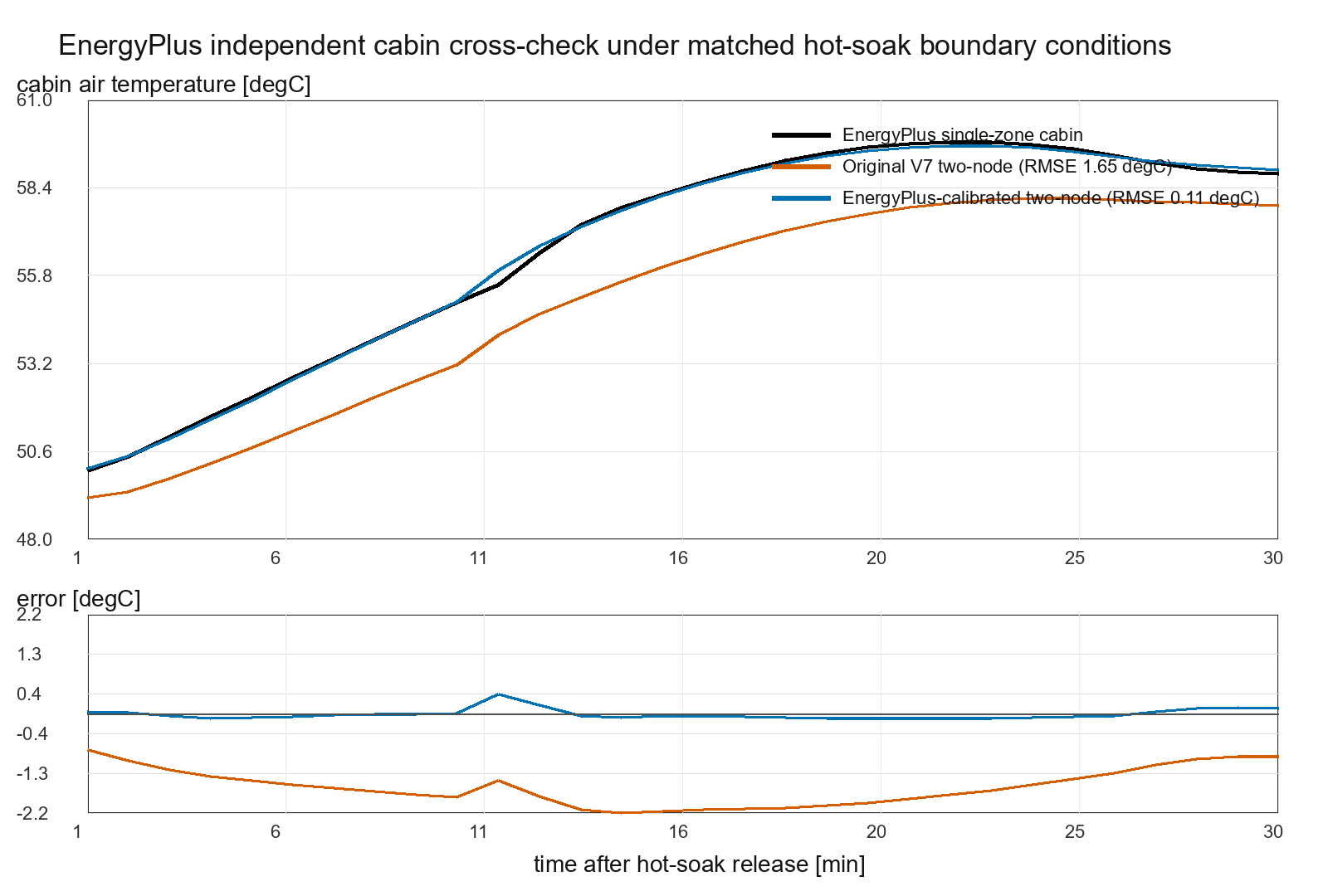}
\caption{Independent cabin cross-check. The two-node cabin (orange) already tracks the EnergyPlus single-zone cabin (black) to 1.65\,$^\circ$C RMSE; bounded physical re-identification (blue) reaches 0.11\,$^\circ$C. Lower panel: pointwise error.}
\label{fig:ep}
\end{figure}

\textbf{Independent cabin cross-check.} Driven by the same hot-soak boundary conditions, an EnergyPlus~25.2.0 single-zone cabin and our two-node cabin agree to \textbf{1.65\,$^\circ$C} RMSE as-is; a light, physically-bounded re-identification of the cabin parameters reduces this to \textbf{0.11\,$^\circ$C} RMSE (Fig.~\ref{fig:ep}), with the identified envelope conductance and solar aperture remaining within their documented physical ranges. This is a cabin thermal-load confirmation only---the full closed-loop high-fidelity evidence remains the OpenModelica FMU.

\subsection{Closed-loop control: six scenarios (shielded Sci-ML MPC vs rule-based)}
\begin{figure}[h]\centering
\includegraphics[width=0.92\linewidth]{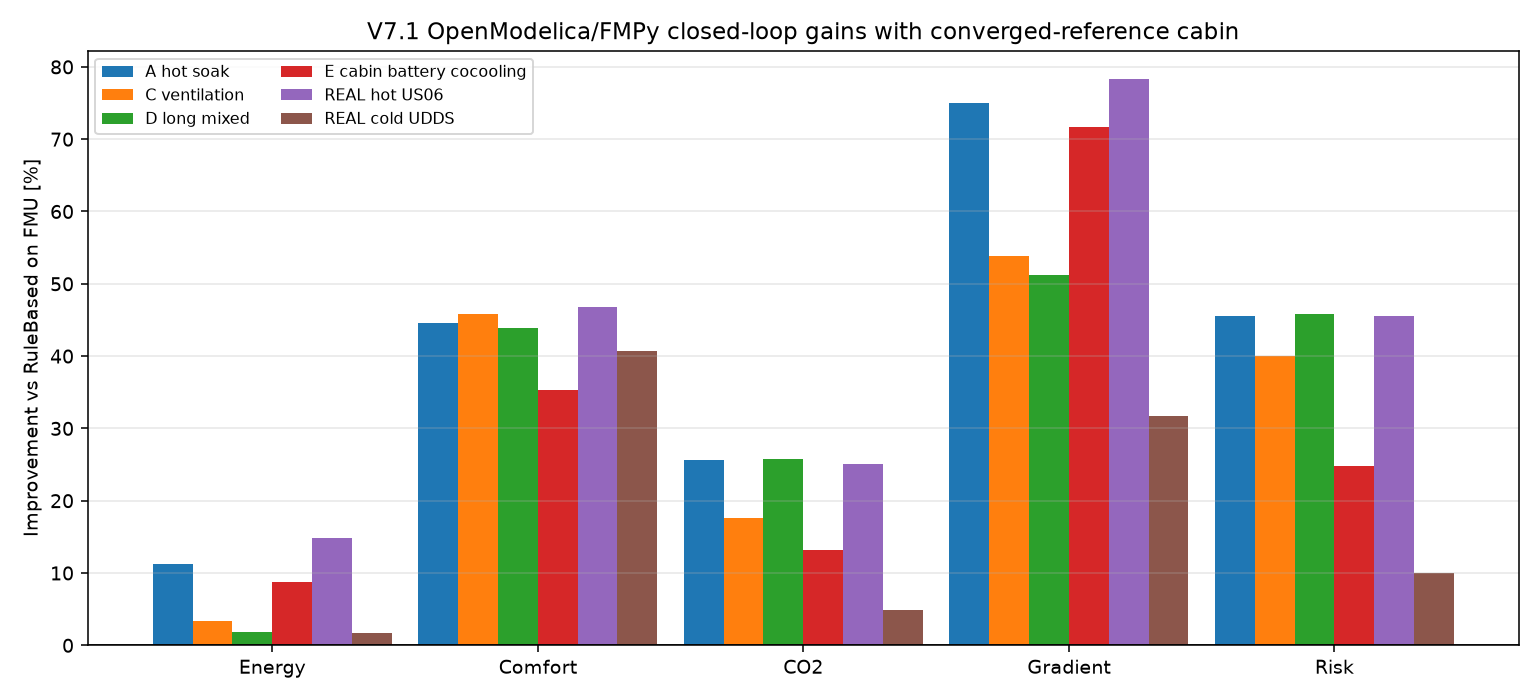}
\caption{Closed-loop improvement of the shielded Sci-ML MPC over the rule-based controller on the independent OpenModelica/FMPy plant, by metric and scenario. Every bar is positive.}
\label{fig:fmu}
\end{figure}

Table~\ref{tab:fmu} and Fig.~\ref{fig:fmu} report the gains on the \emph{independent OpenModelica FMU} plant; all 30 gains are positive and the FMU$\leftrightarrow$Python headline gaps are small (energy $\le0.05\%$). Over 15 seeds on the Python plant, every one of the 30 paired comparisons favours the controller (rank-biserial $=1.0$; Holm-corrected Wilcoxon $p\ll0.01$), and the gains stay positive under $\pm20\%$ perturbation of plant and guard parameters.

\begin{table}[h]\centering\footnotesize\setlength{\tabcolsep}{6pt}
\renewcommand{\arraystretch}{1.15}
\caption{Improvement of the shielded Sci-ML MPC over the rule-based controller (\% better), on the independent OpenModelica 8-node FMU plant.}
\label{tab:fmu}
\begin{tabular}{lccccc}
\toprule
Scenario & Energy & Comfort RMSE & CO\textsubscript{2} max & Battery gradient & Health proxy \\
\midrule
A hot-soak              & 11.3 & 44.6 & 25.6 & 75.0 & 45.6 \\
C ventilation          & 3.4 & 45.9 & 17.6 & 53.8 & 40.0 \\
D long mixed           & 1.8 & 44.0 & 25.8 & 51.3 & 45.9 \\
\textbf{E cabin--battery co-cooling} & \textbf{8.7} & \textbf{35.3} & \textbf{13.1} & \textbf{71.7} & \textbf{24.9} \\
REAL hot US06          & 14.8 & 46.8 & 25.1 & 78.3 & 45.5 \\
REAL cold UDDS         & 1.8 & 40.7 & 4.9 & 31.7 & 10.0 \\
\bottomrule
\end{tabular}
\end{table}

\subsection{Key finding: the cabin--battery co-cooling coupling (scenario E)}
Scenario~E is the physically meaningful coupling point: a $42\,^\circ$C ambient with a hot cabin \emph{and} a hot pack forces the controller to allocate one heat-pump capacity (the \coup{$s_{\mathrm{batt}}$} split) between cabin pull-down and battery protection. Here the rule-based controller either starves the cabin or overheats the pack, whereas the shielded Sci-ML MPC---predicting the coupled dynamics---achieves simultaneous energy ($-8.7\%$), comfort ($-35.3\%$), and battery-gradient ($-71.7\%$) improvement. This is the clearest evidence that the battery and cabin are genuinely coupled through shared capacity and that coordinated, prediction-based allocation is valuable. A second finding: in heating-dominated cold starts the energy headroom is small (rigid heating load), so value shifts to comfort and gradient---a Pareto shift the benchmark exposes rather than hides.

\section{Contributions}
\begin{enumerate}[leftmargin=1.5em]
\item An \textbf{open, reproducible, weather-coupled} whole-vehicle BEV thermal-management benchmark unifying battery health, PMV/PPD comfort, cabin air quality, and energy over six standardised scenarios.
\item A \textbf{real-BMS-anchored} battery thermal core (KU\,Leuven $+$ NASA~POWER; leave-one-vehicle-out validation) and reduced battery/cabin models \textbf{validated against converged higher-fidelity references}, with the cabin additionally cross-checked against EnergyPlus.
\item A \textbf{physics-informed Sci-ML surrogate} that is conservation-exact and dominates black-box and Koopman baselines out-of-distribution.
\item A \textbf{shielded, model-swappable multi-objective MPC} with statistically significant, all-positive gains, cross-validated on an \textbf{independent OpenModelica FMU} closed loop.
\item A fully released package (code, processed data, trained models, parameter-provenance and data-mapping documentation) for reproducible comparison.
\end{enumerate}

\section{Limitations and future work}
Validation is in-silico (reduced-order benchmark $+$ real-data-anchored battery $+$ independent FMU co-simulation $+$ EnergyPlus cabin cross-check); \textbf{hardware-in-the-loop / bench / vehicle tests are the key next step}. Real-data validation currently covers the battery thermal core (cabin/HVAC/ventilation use literature parameters, backstopped by $\pm20\%$ sensitivity and the EnergyPlus cabin cross-check); cross-vehicle evidence is \emph{preliminary} (two vehicles). The aging and ventilation-health terms are comparative proxies, not absolute predictions. A tuned deep-RL (SAC/PPO) comparator and a public archival release (DOI) are planned.

\begin{center}\small\itshape
All quantitative results are produced by the released V7.1 code and pass the internal verification gate (13/13).
\end{center}

\end{document}